\documentclass[12pt]{article}
\usepackage{graphicx,amsmath,amssymb,latexsym}
\newcommand{\beq}{\begin{equation}}
\newcommand{\eeq}{\end{equation}}
\newcommand{\bea}{\begin{eqnarray}}
\newcommand{\eea}{\end{eqnarray}}

\newcommand{\ra}{\right\rangle}
\newcommand{\la}{\left\langle}
\newcommand{\N}{{\cal N}}

\def\lsim{\mathrel{\rlap{\lower4pt\hbox{\hskip1pt$\sim$}}
    \raise1pt\hbox{$<$}}}

\begin{document}

\begin{titlepage}

\begin{flushright}
FIT HE-07-01 \\
MPP-2007-74\\

\end{flushright}
\vspace*{3mm}
\begin{center}
{\bf\Large Holographic heavy-light mesons \\
\vspace*{3mm}
from non-Abelian DBI \\ }

\vspace*{12mm}
Johanna Erdmenger $^a$ \footnote[2]{\tt jke@mppmu.mpg.de}, 
Kazuo Ghoroku $^b$ \footnote[1]{\tt gouroku@dontaku.fit.ac.jp} and
Ingo Kirsch $^c$ \footnote[3]{\tt kirsch@phys.ethz.ch}
\vspace*{2mm}

\vspace*{7mm}

{\em ${}^a$ Max Planck-Institut f\"ur Physik (Werner
  Heisenberg-Institut), F\"ohringer Ring 6, 80805 M\"unchen, Germany}
\vspace*{2mm}\\
{\em ${}^b$Fukuoka Institute of Technology, 
Fukuoka 811-0295, Japan} \\
\vspace*{2mm}
{\em ${}^c$ Institut f\"ur Theoretische Physik, ETH Z\"urich, \\ 
CH-8093 Z\"urich, Switzerland}
\end{center}

\vspace*{20mm}

\begin{abstract}
In the context of gauge/gravity duals with flavor, we examine
heavy-light mesons which involve a heavy and a light quark. For this
purpose we embed two D7 brane probes at different positions into the
gravity background.  We establish the non-Abelian Dirac-Born-Infeld
(DBI) action for these probes, in which the $U(2)$ matrix describing
the embedding is diagonal. The fluctuations of the brane probes
correspond to the mesons. In particular, the off-diagonal elements of
the $U(2)$ fluctuation matrix correspond to the heavy-light mesons,
while the diagonal elements correspond to the light-light and
heavy-heavy mesons, respectively. The heavy-light mesons scale
differently with the 't~Hooft coupling than the mesons involving
quarks of equal mass.  The model describes both scalar and vector
mesons. For different dilaton-deformed gravity backgrounds, we also
calculate the Wilson loop energy, and compare with the meson masses.
\end{abstract}
\end{titlepage}

\section{Introduction}

Recently, based on the gauge/gravity correspondence \cite{MGW}, many
non-perturbative properties of Yang-Mills theories with quarks have
been uncovered in terms of superstring theory
\cite{KK}-\cite{Antonyan:2006pg}.  Flavor quarks in the fundamental
representation of the gauge group are introduced by embedding one or
several probe branes into an appropriate bulk gravity background, in
order to describe large $N$ gauge theories similar to QCD.  Many
successful results have been obtained for the properties of quarks and
their bound states: Mesons spectra have been studied by many authors,
including \cite{KMMW}-\cite{Filev:2007qu}.  The mass spectra of
fermionic operators with fundamental fields (``mesinos'') have been
discussed in \cite{Kirsch:2006he,Heise}.

Up to now, most of these investigations have been devoted to the case
that the D7 branes are embedded at the same place, such that the
flavor group forms $U(N_f)$ for $N_f$ D7 branes.  Effects of the
non-abelian nature of $U(N_f)$ for $N_f>1$ in relation to bulk
instantons have been studied in \cite{Hata:2007mb} and in
\cite{Guralnik:2004ve}-\cite{Arean:2007nh}. Examples of D7 branes 
embedded at different positions occur in studies of meson decay via
string breaking \cite{Peeters:2005fq}-\cite{Bigazzi:2006jt}.

Here we propose a holographic model for heavy-light mesons based on a
non-abelian Dirac-Born-Infeld (DBI) action. In this model, two D7
brane probes are embedded at different positions, such that they
provide different quark mass states depending on the flavor.  It is
possible to choose any mass difference by separating the branes by an
appropriate distance. In terms of this model, we study meson states
with heavy and light quarks, such as mesons with charm or bottom
degrees of freedom for instance. Heavy-light mesons with large spin
have been studied in
\cite{PT,BST}.

 In \cite{EEG}, an effective model for heavy-light
mesons such as the B meson has been proposed which is based on the
Polyakov action. This model gives a qualitative description of
the B and excited B$^*$ states. It also gives rise to a dependence of the
heavy-light meson mass on the 't~Hooft coupling ($\lambda$) of the
form $M_{HL}/m_H = 1 + {\rm const}/\sqrt{\lambda} + {\cal
O}(\lambda^{-1})$, with $M_{HL}$ the heavy-light meson mass and $m_H$
the heavy quark mass (The light quark mass $m_L$ has been set to zero
here.). This differs from the 't~Hooft coupling dependence of the
light-light or heavy-heavy mesons \cite{KMMW}, for which $M \propto
m/\sqrt{\lambda}$.  However the result of \cite{EEG} ensures that the
heavy-light meson mass equals the heavy quark mass in the large
$\lambda$ limit, $M_{HL}=m_H$, as expected from observation and
dimensional analysis.

{Here we present an alternative approach to the same heavy-light meson
mass in terms of a non-Abelian Dirac-Born-Infeld (DBI) action for
two  D7 branes at different positions.}  We use the
non-Abelian DBI action for curved 10d backgrounds which has been
proposed by Myers \cite{My}.  In this action, the world-volume fields
are assigned to $U(N_f)$ matrix-valued functions for $N_f$ D7 branes.
We choose $N_f=2$.  The embedding configuration of the two D7 branes
is determined by the diagonal components of the scalar fields. The
corresponding equation of motion is solved by the profile functions of
two separated branes, one of which corresponds to the heavy and one to
the light quark. The quark masses are given by the boundary values of
the two embedded branes.  The fluctuations of the diagonal elements of
the $2\times 2$ flavor matrices correspond to the light-light and
heavy-heavy mesons, respectively.  On the other hand, the off-diagonal
components of the fluctuations of the fields on the branes are
identified with the heavy-light mesons. We calculate the spectrum of
these heavy-light mesons.

For the $\lambda$ dependence of the heavy-light meson mass, we find
that it is similar to the one observed using the Polyakov action
approach in \cite{EEG}.  A finite contribution to the mass remains in
the limit of $\lambda\to \infty$. This contribution corresponds to the
minimum energy of a classical string connecting two separated D7
branes, and thus is equivalent to the mass obtained from the Polyakov
action.

This $\lambda$ dependence persists if we consider the ${\rm D3} + {\rm
  D(-1)}$ gravity background of~\cite{Liu:1999fc}. In the field theory
dual to this background, a condensate $q \equiv \pi^2 \langle F^2
\rangle$ is switched on. D7 embeddings and chiral symmetry breaking
for a non-supersymmetric version of this background have been studied
in \cite{GY}.  The $\lambda$ dependence is very similar to that 
in the supersymmetric background.

It is instructive to compare the $\lambda$ dependence of the meson
spectra with the $\lambda$ dependence of the tension. For a classical
string stretched between the two D7 brane probes, the string tension
is independent of $\lambda$, in agreement with the heavy-light meson
mass result found in \cite{EEG} as well as in the present paper. For
heavy-light mesons this tension contributes to the meson mass even if
the distance $L$ between the quark and anti-quark in the
four-dimensional boundary space is zero, in which case it contributes
$E=m_H - m_L$ to the Wilson line energy.  For the heavy-heavy and
light-light mesons, the string tension scales as
$m_q^2/\sqrt{\lambda}$ for small $L$ \cite{KMMW,GY}. At large $L$,
when the dual gauge theory is in the quark confinement phase, there is a
long range linear potential for all the mesons considered. For the
heavy-heavy and light-light mesons this was found in~\cite{GY}.

Our model allows to describe both scalar and vector mesons. In this
respect, it goes beyond the effective model of \cite{EEG}. For the
supersymmetric ${\rm D3} + {\rm D(-1)}$ background of
\cite{Liu:1999fc}, we find that the vector and scalar meson masses
differ, due to the different dependence of the fluctuation equations
on the non-trivial dilaton. Note that when adding D7 probes to the
${\rm D3} + {\rm D(-1)}$ background, supersymmetry is broken to ${\cal
N}=1$ and the vector and scalar mesons are not in the same multiplet
any more. For very large heavy quark mass, ${\cal N}=2$ supersymmetry
is restored, and the vector and scalar meson masses become degenerate
again.  This is consistent with the phenomenological fact from
heavy-quark theory that spin effects are suppressed by powers of the
inverse heavy quark mass.  Of course, here this is due to ${\cal N}=2$
supersymmetry restoration. We leave an investigation of this mechanism
for non-supersymmetric backgrounds to the future.

\vspace{.2cm} 
Our paper is organized as follows.  In section 2, we give the
non-Abelian DBI action and the D7 brane embedding model is proposed.
In section 3, the number of D7 branes is restricted to two ($N_f=2$)
and the action is expanded by fluctuations to see the meson spectra,
which are shown in the section 4 for the case of HL mesons. In section 5,
the potential between quark and anti-quark is given through Wilson loop. The
summary is given in the final section.

\newpage
\section{Embedding of D7 branes}

\subsection{Non-Abelian Dirac-Born-Infeld action}

We start from the non-Abelian Dirac-Born-Infeld action proposed by
Myers in \cite{My}. This action describes the dynamics of $N_f$
D$p$-branes in a background with metric $G_{mn}$ and is given by
\begin{align} 
S_{N_f} = - \tau_p \int d^{p+1} \xi  e^{-\phi} {\,\rm STr} \left(
\sqrt{-\det ( P[G_{rs} + G_{ra} (Q^{-1} -\delta)^{ab} G_{sb}] 
+ T^{-1} F_{rs})}
\sqrt{\det Q^a{}_b} \right) \,, \label{two-brane}
\end{align}
where the matrix $Q^a{}_b$ is defined by
\beq 
 Q^a{}_b = \delta^a{}_b + i T [X^a,X^c] G_{cb} \,
\eeq 
where $T^{-1} =2\pi\alpha '$, and $X^a$ are the coordinates transverse 
to the stack of branes, which now take values in a  $U(N_f)$ algebra.
The symbol ${\rm STr}$ denotes the
symmetrized trace ${\rm STr} (A_1 ...A_n) \equiv \frac{1}{ n!} {\rm
  Tr}(A_1...A_n $ + all permutations) and is needed to avoid the
ambiguity of the ordering of the expansion of 
all fields in the DBI action.
\cite{Tsey}.

In our convention, $r,s= 0,1,...,p$ and $a,b= p+1, ..., 9$ label the
world-volume directions and the directions transverse to the D$p$-branes,
respectively; $m,n=0,1, \cdots, 9$ are the 10d spacetime indices.
$P[a_{rs}]$~denotes the pull-back of a 10d tensor $a_{mn}$ to the
world-volume of the branes. A peculiarity of the non-Abelian DBI
action is that the pull-back matrix is given by the covariant
derivative
\beq  
D_r X^a=\partial_{r}X_a + i [A_r, X^a]\ ,
\eeq
with partial derivatives $\partial_r \equiv \partial/\partial \xi^r$, 
non-Abelian world-volume gauge field $A_r$ and transverse
coordinates $X^a$. $F_{rs}$ is the corresponding world-volume field
strength. 

\vspace{.3cm}
For diagonal brane embeddings the commutator $[X^a,X^b]$ is
small, as can be seen as follows. Set $X^a$ as
$X^a=\bar{X}^a+\phi^a$, where $\phi^a$ denotes small quantum
fluctuations around a {\em diagonal} embedding matrix $\bar X^a$.
Then we have
\bea
[X^a,X^b]=[\bar X^a,\bar X^b]+ [\bar X^a, \phi^b]
         +[\phi^a, \bar X^b]+ [\phi^a,\phi^b]\,,
\eea
where the first term $[\bar X^a,\bar X^b]$ vanishes and the
remaining terms are small. Next, assuming also a diagonal metric
$G_{mn}$ and employing the approximation $(Q^{-1} -\delta)^{ab}
\approx -iT [X^a,X^b]$, we rewrite the pull-back in the 
action~(\ref{two-brane})~as
\bea 
P[G_{rs} + G_{ra} (Q^{-1} -\delta)^{ab} G_{sb}]
\,\approx\, G_{rs} + D_r X^a D_s X^b \left(
G_{ab} - i T [X^c, X^d]G_{ac}G_{bd} \right) \,. 
\eea 
Then, the action (\ref{two-brane}) is expanded in powers of $[X^a,X^b]$ up to 
$O\!\left(X^4\right)$ as 
$$ 
S_{N_f}= 
  \tau_p   \int d^{p+1} \xi \
 e^{-\Phi}{\rm STr}  \left\{  
  \sqrt {-\det 
({G}_{rs}  + G_{ab}D_r X^a D_s X^b
 +   T^{-1} F_{rs})}    \right. 
$$
\beq
 \times  \left.\left(1-{1\over 4}\left(TG_{ac} [X^c,X^b]\right)^2\right) \ \right\} \ .
\label{two-brane-2}
\eeq
The factor in the second line descends from the expansion of
$\sqrt{\det Q^a{}_b}$. For a flat spacetime background
$G_{mn}=\eta_{mn}$, the action (\ref{two-brane-2}) agrees with that
found in \cite{Tsey}.

\subsection{Diagonal embedding ansatz in non-Abelian DBI action}

The non-Abelian DBI action will now be used to find the embedding of
$N_f$ probe D7 branes in different gravity backgrounds. The embedding
profiles correspond to the classical solutions for the scalar fields
in the D7 brane action.  In our case, the scalar fields $X^a$ are
$U(N_f)$ matrix valued functions which makes it difficult to obtain a
general form of the profile functions. In order to simplify the
problem, we use the diagonal ansatz
\beq \label{diagonal}
 X^a={\rm diag}(w_1^a, \cdots, w_{N_f}^a) \, ,
\eeq
thereby setting all off-diagonal components to zero.  Here each of the
functions $w_i^a$ corresponds to one of the $N_f$ D7 branes. -- It would
be an interesting problem to also include the off-diagonal components
of $X^a$ and to solve the corresponding embedding equations. For
example, for a non-trivial world-volume gauge field $F_{rs}$ the
off-diagonal components provide a BI-on configuration which connects
two branes \cite{CM}. We postpone the discussion of such
configurations to the future.

\vspace{.3cm} 
The quark mass for each flavor is given by the asymptotic value of
$w_i^a$ in the ultraviolet limit. They are the integration constants
and given by hand as parameters of the theory.\footnote{The asymptotic
values of (possible) off-diagonal components, say $w_{ij}^a$,
represent the mass-mixing of different quark flavors $i$ and $j$. We
neglect these here.}  The equations of motion for the $w_i^a$ are
obtained from the action
\begin{align}
 S_{N_f} &= 
  \tau_7   \int d^{8} \xi \  e^{-\Phi}{\rm STr}  \left(  
  \sqrt {-\det ({G}_{rs}  + G_{ab}\partial_r w_i^a \partial_s w_i^b )}
  \right) \nonumber \\
 &= \tau_7   \int d^{8} \xi \  e^{-\Phi}\sum_{i=1}^{N_f}  
  \sqrt {-\det ({G}_{rs}+ G_{ab}\partial_r w_i^a \partial_s w_i^b )}\, 
  \label{two-brane-3}
\end{align}
which is Eq.~(\ref{two-brane-2}) for the embedding (\ref{diagonal})
and $p=7$. The essential point is here that for the diagonal ansatz
(\ref{diagonal}), we obtain $N_f$ decoupled equations of motion for
the $w_i^a$ such that the embeddings of each of the probe branes is
independent of the other. In other words, for diagonal embeddings the
non-Abelian DBI action reduces to the sum of $N_f$ abelian DBI
actions.\footnote{This is only true, if we ignore the fluctuations
around the brane embeddings. As we will see in Sec.~3, the non-Abelian
DBI action also describes fluctuations of strings stretched in between
two different branes.}

If there are additional fluxes in the background, the action
(\ref{two-brane-3}) must be supplemented by appropriate Wess-Zumino
terms.
  
\subsection{Specific gravity background}

\vspace{.3cm}
Let us now find the embedding functions $w \equiv w_i$ ($i=1,...,N_f$)
of the probe branes for some phenomenologically interesting
supergravity backgrounds.

As a specific supergravity background, we consider the following 10d
background in string frame given by a non-trivial dilaton $\Phi$ and
axion $\chi$ \cite{Liu:1999fc,GY},
\beq
ds^2_{10}= e^{\Phi/2}
\left(
\frac{r^2}{R^2}A^2(r)\eta_{\mu\nu}dx^\mu dx^\nu +
\frac{R^2}{r^2} dr^2+R^2 d\Omega_5^2 \right) \ .
\label{non-SUSY-sol}
\eeq
Here two typical solutions are considered. One is the
supersymmetric solution
\beq 
A=1, \quad e^\Phi= 
1+\frac{q}{r^4} \ , \quad \chi=-e^{-\Phi}+\chi_0 \ ,
\label{dilaton} 
\eeq
and the other is non-supersymmetric and given by
\beq
 A(r)=\left(1-(\frac{r_0}{r})^8 \right)^{1/4}, \quad \chi=0 \ , \quad
 e^{\Phi}=
 \left(\frac{(r/r_0)^4+1}{(r/r_0)^4-1} \right)^{\sqrt{3/2}} \ .
\label{dilaton-2}
\eeq 
The first solution is dual to $\N=2$ super Yang-Mills theory with
gauge condensate $q$ and is chirally symmetric. Both supersymmetry and
chiral symmetry are broken for the second solution \cite{GY}.

\vspace{.3cm}
In order to obtain the induced metric on the D7 world-volume, we rewrite
the six-dimensional part of the metric (\ref{non-SUSY-sol})
in the form
\beq
 \frac{R^2}{r^2} dr^2+R^2 d\Omega_5^2
 =\frac{R^2}{r^2}\left(d\rho^2+\rho^2d\Omega_3^2+(dX^8)^2+(dX^9)^2
\right)\ ,
\eeq
where $r^2=\rho^2+(X^8)^2+(X^9)^2$. Due to the rotational invariance
in the $X^8-X^9$ plane, we may set $X^8=0$ and $X^9=w(\rho)$ without
loss of generality. Then the induced metric on the D7 brane is given
by
\beq
ds^2_8=e^{\Phi/2}
\left\{
{r^2 \over R^2}A^2\eta_{\mu\nu}dx^\mu dx^\nu +\right.
\left.\frac{R^2}{r^2}\left((1+(\partial_{\rho}w)^2)d\rho^2
+\rho^2d\Omega_3^2\right)
 \right\} \ .
\label{D7-metric}
\eeq

\vspace{.3cm}
In static gauge the action for the D7 probe is given by 
\beq
S_{\rm D7} =  S_{DBI} + S_{WZ} = -\tau_7~\int d^8\xi  \sqrt{\epsilon_3}\rho^3
\left(A^4 e^{\Phi}\sqrt{ 1 + (w')^2 }-C_8\right)
\ ,
\label{D7-action-2}
\eeq
where $C_8=q/r^4=e^{\Phi}-1$ denotes the Wess-Zumino term coming from
$A_8$, the Hodge dual of the axion \cite{GY}. The Wess-Zumino term
is only required for the supersymmetric background (\ref{dilaton}).  
The equation of motion for the
embedding function $w(\rho)$ is 
\bea
 && - {w\over \rho+w~w'}
  \sqrt{1+(w')^2}(\Phi+4\log A)'
\nonumber\\
 && 
\quad +{1\over \sqrt{1+(w')^2}}
\left[w'\left({3\over \rho}+(\Phi+4\log A)'\right)
+ {w''\over 1+(w')^2}\right]
   =0 
  \label{qeq}
\eea
for the non-supersymmetric case (\ref{dilaton-2}), and
\beq 
{w\over \rho+w~w'}
   \Phi'\left[1-\sqrt{1+(w')^2}~\right]+{1\over \sqrt{1+(w')^2}}
\left[w'\left({3\over \rho}+\Phi'\right)
+ {w''\over 1+(w')^2}\right]
   =0 
  \label{qeq2}
\eeq
for the supersymmetric case (\ref{dilaton}). Here the prime denotes
the derivative with respect to $\rho$.

In deriving these equations, we have to take into account that
$r=\sqrt{\rho^2+w(\rho)^2}$. We therefore have to extract the variation
of $w$ also from the functions of $A(r)$ and $\Phi(r)$. For example,
the variation of $A(r)$ with respect to $w$ is obtained as
$$ \delta A(r)={\partial r^2\over \partial w}\partial_{r^2}A(r) \delta w+\cdots
       ={w\over \rho+w\partial_{\rho}w}\partial_{\rho}A\delta w+\cdots
$$
Here, the expression after the second equality sign shows the change
of variable from $r$ to $\rho$ in the derivative.  The prefactor of
the first term of (\ref{qeq}) and (\ref{qeq2}) originates from this
variable changing procedure.

\vspace{.3cm} 
Solving the above equation for $w$, we find the profile functions of
the D7 brane embedded at the separated places and then we find
simultaneously the quark properties, the quark mass $m_q$ and the
chiral condensate $\la\bar{\Psi}\Psi\ra$, where $\Psi$ denotes the
quark field. The details of the solutions are shown in \cite{GY}.
 
\section{D7-brane fluctuations} \label{sec3}

In this section we derive the actions of the scalar and vector D7
brane fluctuations dual to heavy-heavy, light-light and heavy-light
mesons. These actions will be used in the next section to find the
fluctuation spectrum in the backgrounds of the type
(\ref{non-SUSY-sol}).

For this, we return to the brane action~(\ref{two-brane-2}).  At this
stage, we restrict to the case of $N_f=2$ flavors or two D7 branes
such that the scalar and vector fields in the non-Abelian DBI action
are represented by $2\times 2$-matrices. For the classical embedding, 
we choose
the diagonal configuration given by
\beq
 \bar{X}^8=0\,,\qquad
 \bar{X}^9=\left(\begin{array}{cc}
            w_1&0\\
            0&w_2 \end{array}\right)\, .
\eeq
In terms of the Pauli matrices
\beq
 \tau^0={1\over 2}
        \left(\begin{array}{rr}
            1&0\\
            0&1 \end{array}\right)\, ,\,\,
\tau^1={1\over 2}
       \left(\begin{array}{rr}
            0&1\\
            1&0 \end{array}\right)\, ,\,\,
\tau^2={1\over 2}
       \left(\begin{array}{rr}
            0&-i\\
            i&0 \end{array}\right)\, ,\,\,
\tau^3={1\over 2}
       \left(\begin{array}{rr}
            1&0\\
            0&-1 \end{array}\right)\,, 
\eeq
$\bar X^9$ can be rewritten as
\beq
 \bar{X}^9=w\tau_0+v\tau_3\,, \quad w_1=(w+v)/2\,, \quad w_2=(w-v)/2\,,
 \label{profile}
\eeq 
where $v=w_1-w_2$. The asymptotic boundary values of 
$w_1$ and $w_2$ correspond to the heavy and light quark masses, respectively.  
When $v=0$, the two branes are at the same place, $w_1=w_2=w$,
corresponding to a $U(2)$ flavor symmetry. For $v \neq 0$ this flavor
symmetry is explicitly broken.

The scalar and gauge field fluctuations are taken to be of the form $(a=8, 9)$ 
\beq
X^9=\bar{X}^9+\phi^9\,, \quad X^8=\phi^8\,, \label{expansions} 
\eeq
\beq \phi^a=\phi^a_0\tau^0+\phi^a_i\tau^i\, , \quad
A^r=A^r_0\tau^0+A^r_i\tau^i\, , \label{flucs}  
\eeq 
and can be written as 
\beq
\phi^a = \left(\begin{array}{cc}
    \phi^a_+ & \phi^a_{12}\\
    \phi^a_{21}& \phi^a_- \end{array}\right) \,,
\eeq 
similarly $A^r$.  The diagonal elements $\phi^a_\pm=\phi^a_0 \pm
\phi^a_3$ describe fluctuations of each brane and are dual to the
heavy-heavy and light-light mesons. On the other hand, the
off-diagonal elements $\phi^a_{12}=\phi^a_1-i\phi^a_2$ and
$\phi^a_{21} = \phi^a_1+i\phi^a_2$ correspond to fluctuations of
strings stretched between the two branes and are dual to the
heavy-light mesons. The mass of this last type of fluctuations will
depend on $v$. -- A similar structure emerges also for gauge field
fluctuations $A_r$, as discussed below.

\vspace{.5cm}
These meson mass spectra are obtained by solving the linearized
equation of motions for the field fluctuations. Using the expansions
(\ref{expansions}) and assuming small fluctuations $\phi^a$ and $A_r$,
the action (\ref{two-brane-2}) is rewritten as
\beq
 S_{N_f=2}= 
  \tau_7   \int d^{8} \xi ~
  {\rm STr}  \left\{  e^{-\Phi}
   \sqrt {-\det 
   ({a}_{rs})}\left(1+G_{88}G_{99}{1\over 8}\left((\phi_1^8)^2
   +(\phi_2^8)^2\right)v^2\right) \ \right\} \ , 
\label{two-brane-32}
\eeq
where
\beq
   {a}_{rs} \equiv {G}_{rs} + G_{ab}D_r X^a D_s X^b 
   + T^{-1} F_{rs}\, .  \label{DB-bg}
\eeq
and $\phi_1^8$ and $\phi_2^8$ as defined in (\ref{flucs}). The other
components of $\phi^8$ and $\phi^9$ do not appear explicitly, but
contribute to ${a}_{rs}$ in (\ref{two-brane-32}). 

When evaluating (\ref{two-brane-32}), we have to take into account
that the radial coordinate $r$, which occurs in $G_{rs}, G_{ab}$ and
$\Phi$ on the brane, is a matrix-valued function. Including the
fluctuations, $r$ is of the form
\beq
 r^2=\rho^2+(X^8)^2+(X^9)^2
    =\rho^2+(w\tau_0+v\tau_3+\phi^9)^2+(\phi^8)^2 \, \label{r}
\eeq
and, for instance, $G_{rs}$ denotes the matrix
\begin{equation}
G_{rs} \leadsto \left(\begin{array}{cc}
            G_{rs}\vert_{r=r_{11}} & G_{rs}\vert_{r=r_{12}}\\
            G_{rs}\vert_{r=r_{21}} & G_{rs} \vert_{r=r_{22}} 
            \end{array}\right),
\end{equation}
where $r_{ij}$ ($i,j=1,2$) are the matrix elements of $r$.

It is convenient to write ${a}_{rs}$ in (\ref{two-brane-32}) as
\beq
   {a}_{rs}=\bar{{a}}_{rs}+\delta {a}_{rs} \,,
\eeq
where
\beq \label{baras}
\bar{a}_{rs} = G_{rs}+G_{99}\partial_r \bar{X}^9 \partial_s\bar{X}^9
\,.
\eeq
The explicit fluctuation dependent part $\delta {a}_{rs}$ is expanded
in terms of the power series of the fluctuations as
\beq
 \delta {a}_{rs}={a}_{rs}^{(1)}+{a}_{rs}^{(2)}+ ...
 \,,
\eeq
where
\begin{align}
 {a}_{rs}^{(1)}&=G_{99}\left(\partial_r \bar{X}^9\partial_s \phi^9
     +\partial_r \phi^9\partial_s \bar{X}^9
     -i[A_r, \bar{X}^9]\partial_s\bar{X}^9
        -i\partial_r\bar{X}^9[A_s, \bar{X}^9]\right)
+ T^{-1}F_{rs}\, , \label{1st-fl} \\
 {a}_{rs}^{(2)}&=G_{99}\left(-i[A_r, \bar{X}^9]\partial_s\phi^9
    -i\partial_r\phi^9[A_s, \bar{X}^9]+\partial_r \phi^9\partial_s \phi^9
     -[A_r, \bar{X}^9][A_s, \bar{X}^9]\right. \nonumber\\
& \hspace{0.5cm}
     \left.-i[A_r, \phi^9]\partial_s\bar{X}^9
    -i\partial_r\bar{X}^9[A_s, \phi^9] \right)
       +G_{88}\partial_r \phi^8\partial_s \phi^8\, .
\label{2nd-fl}
\end{align}
Moreover, since the metric depends on $r$ given by (\ref{r}),
$\bar{a}_{rs}$ in (\ref{baras}) still includes (implicitly) the matrix-valued
quantum fluctuations $\phi^8$ and $\phi^9$.

Then the action (\ref{two-brane-2}) is expanded up to quadratic order
in the fluctuations,
$$ 
S_{N_f=2}= 
  \tau_7   \int d^{8} \xi \
 {\rm STr}  \left\{e^{-\Phi}  
  \sqrt {-\det 
(\bar{a}_{rs})}\left( 1+{1\over 2}{\rm tr}_{rs}(\bar{a}^{-1}{a}^{(1)})
+{1\over 8}\left({\rm tr}_{rs}(\bar{a}^{-1}{a}^{(1)})\right)^2
    \right. \right. 
$$
\beq
\mbox{}\quad\,\left. -{1\over 4}{\rm tr}_{rs}\left((\bar{a}^{-1}{a}^{(1)})^2\right)
+{1\over 2}{\rm tr}_{rs}(\bar{a}^{-1}{a}^{(2)})
\left. -{1\over 8}G_{88}G_{99}\left((\phi_1^8)^2+(\phi_2^8)^2\right)v^2
+\cdots\right) \ \right\} \ . 
\label{two-brane-42}
\eeq

\subsubsection*{Lagrangian for scalar fluctuations}

From (\ref{2nd-fl}), we see that the fluctuations $\phi^9$ and
$A_r$ are mixing, while $\phi^8$ does not mix with any other field.
For simplicity, we consider only the $\phi^8$ fluctuations. When 
evaluating the symmetrized trace in (\ref{two-brane-42}) it has to be
kept in mind that the diagonal flavor matrix elements of the embedding
have to be evaluated at $w_1(\rho)$ and $w_2(\rho)$, respectively. 
To quadratic order the Lagrangian for the $\phi^8$ fluctuations reads 
\begin{align}
 {\cal L}_{\phi^8}^{(2)}&={1\over 4}\partial_{r^2}\bar{F}\vert_{w_1} \left(
   (\phi^8_{+})^2+(\phi^8_1)^2+(\phi^8_2)^2\right) 
 +  {1\over 4}\partial_{r^2}\bar{F}\vert_{w_2} \left(
   (\phi^8_{-})^2+(\phi^8_1)^2+(\phi^8_2)^2\right) \nonumber\\
 & \quad+{1\over 8} 
   (\bar{F}~G_{88}\bar{a}^{rs})\vert_{w_1}
   \partial_r \phi^8_{+}\partial_s\phi^8_{+}
   +{1\over 8} (\bar{F}~G_{88}\bar{a}^{rs})\vert_{w_2}
   \partial_r \phi^8_{-}\partial_s\phi^8_{-}       \nonumber\\
 & \quad+{1\over 8}\left( (\bar{F}~G_{88}\bar{a}^{rs})\vert_{w_1}
   +(\bar{F}~G_{88}\bar{a}^{rs})\vert_{w_2}\right)
   \left(\partial_r \phi^8_{1}\partial_s\phi^8_{1}
   +\partial_r \phi^8_{2}\partial_s\phi^8_{2}\right)\nonumber\\
 & \quad-{v^2\over 8}\left( (\bar{F}~G_{88}G_{99})\vert_{w_1}
   +(\bar{F}~G_{88}G_{99})\vert_{w_2}\right)
   \left((\phi^8_1)^2+(\phi^8_2)^2\right) \,, \label{phi12}
\end{align}
where 
\beq
 \bar{F}=e^{-\Phi}\sqrt{-\det \bar{a}_{rs}} \,.
\eeq
In the above equation (\ref{phi12}), the notation $K(r)\vert_{w_i}$
means that $r$ in any function $K(r)$ is replaced by $\bar
r_i=\sqrt{\rho^2+w_i^2}$, $K(r)\vert_{w_i}=K(\sqrt{\rho^2+w_i^2})$.
Moreover, the first two terms of~(\ref{phi12}) are derived from the
matrix-valued coordinate $r$, which includes the fluctuations $\phi^8$
and $\phi^9$. The first two terms of (\ref{phi12}) are essential for
finding the correct spectrum. In particular, in the non-supersymmetric
case these terms are crucial for finding the Nambu-Goldstone bosons of
chiral symmetry breaking.

From the Lagrangian (\ref{phi12}) we obtain the standard equations of
motion for the heavy-heavy and light-light modes $\phi^8_\pm$ as well
as a new equation for the heavy-light fluctuations. Note that the mass
spectra of $\phi_{1,2}^8$ depend on both profile functions
$w_{1,2}(\rho)$.  -- For the details of the meson spectrum in the
single brane case see \cite{BGN}.

\subsubsection*{Lagrangian for vector fluctuations}

For the vector meson, we find a similar separation of the modes as for
the scalars. However there is a mixing of the $A_r^{1,2}$ and $\phi^9$
modes, as we now discuss.  For the case of constant $w^i$, we can see
that this mixing can be removed by the gauge transformation of
$A_r^{1,2}$. By taking
\bea
 A_r^{1}&=&\tilde{A}_r^{1}+{1\over v}\partial_r\phi_{2}^9 \, , \nonumber \\
 A_r^{2}&=&\tilde{A}_r^{2}-{1\over v}\partial_r\phi_{1}^9\, ,
\eea
the mixing part in (\ref{2nd-fl}) is written as
$$a^{(2)}_{\tilde{A},\phi^9}=G_{99}\left(-i[A_r, \bar{X}^9]\partial_s\phi^9
    -i\partial_r\phi^9[A_s, \bar{X}^9]+\partial_r \phi^9\partial_s \phi^9
     -2[A_r, \bar{X}^9][A_s, \bar{X}^9]\right) . $$
Again, when explicitly writing out the flavor matrix, we have a diagonal form
\beq
 a^{(2)}_{\tilde{A},\phi^9}=G_{99}\left\{\left(\begin{array}{rr}
            \partial_r \phi^9_{+}\partial_s\phi^9_{+}&~~~{}\\
            ~~~{}&\partial_r \phi^9_{-}\partial_s\phi^9_{-} \end{array}\right)
         +\left(\begin{array}{rr}
            1&0\\
            0&1 \end{array}\right)
          {v^2\over 4}\left(\tilde{A}_r^{1}\tilde{A}_s^{1}
            +\tilde{A}_r^{2}\tilde{A}_s^{2}\right)
\right\}\, , \label{A-fl-2}
\eeq
where the top left entry is evaluated at $w_1$ and the bottom right at
$w_2$.  From (\ref{A-fl-2}) we see that the kinetic terms of
$\phi^9_{1,2}$ are eliminated by changing $A_r^{1,2}$ to the new
variables $\tilde{A}_r^{1,2}$. On the other hand, due to gauge
invariance, the kinetic term of $A_r$ does not give rise to any new
kinetic terms for $\phi^9_{1,2}$.  Instead, new mass terms are
generated for $A_r^{1,2}$ as shown above. The two scalar components
$\phi^9_{1,2}$ are gauged away to produce the longitudinal component
of the vector $A_r^{1,2}$.

This is the well-known Higgs mechanism, with $X^9$ the Higgs scalar
and the $A_r^a$. the $SU(2)$ gauge fields.  $X^8$ is not involved in
this Higgs mechanism of the gauge symmetry breaking. However $X^8$ is
associated with the Nambu-Goldstone mode of the geometrical $U(1)_A$
chiral symmetry breaking. These two symmetry breaking mechanisms are
not related to each other.

\vspace{.3cm}
For constant $w_{1,2}$, the vector meson part of (\ref{two-brane-42})
is given by
\beq
 {\cal L}^{(2)}_{\tilde{A}}={\rm STr}  \left\{  
  e^{-\Phi}\sqrt{-\det \bar{a}_{rs}}\left(-{1\over 4}{\rm tr}_{rs}
\left((\bar{a}^{-1}{a}^{(1)}_{\tilde{A}})^2\right)
+{1\over 2}{\rm tr}_{rs}(\bar{a}^{-1}{a}^{(2)}_{\tilde{A}})\right)
\right\} \ , \label{A2}
\eeq
where in flavor space
\begin{gather}
  {a}^{(1)}_{\tilde{A}}=F_{rs}(\tilde{A}) \left(
\begin{array}{rr} 1&0\\ 0&1 \end{array}\right) \, ,
\end{gather}
\beq
{a}^{(2)}_{\tilde{A}}=G_{99}\left(\begin{array}{rr}
            1&0\\
            0&1 \end{array}\right)
          {v^2\over 4}\left(\tilde{A}_r^{1}\tilde{A}_s^{1}
            +\tilde{A}_r^{2}\tilde{A}_s^{2}\right)  \, . \label{A3}
\eeq
Then we obtain
\begin{align}
 {\cal L}_{\tilde{A}}^{(2)}
  &={1\over 4}
   (\bar{F}\bar{a}^{rp}\bar{a}^{sq})\vert_{w_1} \tilde{F}_{rs}^+\tilde{F}_{pq}^+
   +{1\over 4} (\bar{F}\bar{a}^{rp}\bar{a}^{sq}) \vert_{w_2}
   \tilde{F}_{rs}^-\tilde{F}_{pq}^- \nonumber\\
  &\quad+{1\over 4}\left( (\bar{F}\bar{a}^{rp}\bar{a}^{sq})\vert_{w_1}
   +(\bar{F}\bar{a}^{rp}\bar{a}^{sq})\vert_{w_2}\right)
   \left(\tilde{F}_{rs}^1\tilde{F}_{pq}^1+\tilde{F}_{rs}^2\tilde{F}_{pq}^2\right)
   \nonumber\\
  &\quad+{v^2\over 8}\left( (\bar{F}~G_{99}\bar{a}^{rs})\vert_{w_1}
   +(\bar{F}~G_{99}\bar{a}^{rs})\vert_{w_2}\right)
   \left(\tilde{A}_r^{1}\tilde{A}_s^{1}
   +\tilde{A}_r^{2}\tilde{A}_s^{2}\right) \, , \label{A2-2}
\end{align}
where 
\beq
 \tilde{F}^{\pm}_{rs}=\tilde{F}^{0}_{rs}\pm \tilde{F}^{3}_{rs} \, .
\eeq
Again the dependence on the position of the two distinct branes,
characterized by the two profile functions $w_i(\rho)$, $i=1,2$, is
stated explicitly as in (\ref{phi12}) above.

\vspace{.3cm}
We give a brief summary of this section.  (i) For heavy-heavy or
light-light mesons, the modes of $\phi^a_{0,3}$ and $A^r_{0,3}$
recombine into $\phi^a_{\pm}$ and $A^r_{\pm}$, and their spectra
coincide with the one obtained from one individual probe brane.  (ii)
For heavy-light mesons, only $\phi^8_{1,2}$ and $A^r_{1,2}$ remain as
independent fluctuation variables.  $\phi^9_{1,2}$ are gauged away and
$A^r_{1,2}$ become massive as a result of the Higgs mechanism.  (iii)
The mesons corresponding to $\phi^8_{1,2}$ have a mass which depends
on both brane embeddings $w_1$ and $w_2$.  These modes are not
affected by the above Higgs mechanism.

\newpage
\section{Mass spectra of heavy-light mesons}

In this section we examine the mass spectra of heavy-light (HL) mesons
in the two background solutions.

\subsection{Supersymmetric case}

First, we consider the supersymmetric background 
solution given by (\ref{dilaton}).

\subsubsection{Scalar fluctuations}

For the supersymmetric solution (\ref{dilaton}), the equation of motion
for the fluctuations $\phi^8_{1,2}$ dual to HL mesons is obtained from 
(\ref{phi12}). We find
\beq
  \left(\partial_{\rho}^2+{3\over \rho}\partial_{\rho}
            -{l(l+2)\over \rho^2}+
            {M_{1}+M_{2}\over e^{\Phi(r_1)}+e^{\Phi(r_2)}}\right)
          \phi=0  \label{H-L-phi} \,,
\eeq
\beq
    M_{i}=e^{\Phi(r_i)}\left(\partial_{\rho}\Phi(r_i)\partial_{\rho}
              +{M^2-v^2e^{\Phi(r_i)}\over r^4_i}R^4\right) \,,
\eeq
where $r_i^2 =\rho^2+w_i^2$ $(i=1,2)$ and $\phi^8_{1,2}$ are denoted
by $\phi$. Here $-{l(l+2)}$ is the eigenvalue of the Laplace operator
on $S^3$.

In order to study the qualitative behavior of the spectrum, we first
consider the non-confining case $q=0$.  In this case, the dilaton is
trivial, $\Phi=0$, and Eq.~(\ref{H-L-phi}) simplifies to \beq
\left(\partial_{\rho}^2+{3\over \rho}\partial_{\rho} -{l(l+2)\over
    \rho^2}+{M^2-v^2\over 2}\left( \left({R^2\over
        \rho^2+w_1^2}\right)^2+ \left({R^2\over
        \rho^2+w_2^2}\right)^2\right) \right) \phi=0 \,.
\label{H-L-phi-2} 
\eeq 
The mass of the fluctuations $M$ is interpreted as the heavy-light
meson mass in the dual gauge theory.  Since $w_1$ and $w_2$ are mixed
in a complicated way, we must solve this equation numerically.

Before proceeding to the numerical solution, we consider two special
cases in which the spectrum can be determined analytically.  For
$w_1=w_2=w$, we get $v=0$ and the equation reduces to the one given by
Kruczenski {\em et al} \cite{KMMW} which can be solved analytically.
In this case, the meson masses can be expressed in terms of the quark
mass $m=w/(2\pi \alpha')$ and the 't~Hooft coupling $\lambda=R^4/4\pi
\alpha'^2$ by \beq M^2=4\pi\frac{m^2}{\lambda} \,(n+l+1)(n+l+2) \,,
\label{massHH} \eeq where $n$ denotes the node number of the eigen
functions for $l=0$ and and $l$ represents the angular momentum of
$S^3$ in the world-volume of the D7 brane.  This spectrum represents
therefore the one for the heavy-heavy and light-light mesons.

The other regime in which (\ref{H-L-phi-2}) can be solved analytically
corresponds to a heavy-light meson with a very heavy quark, $w_2 \gg
w_1$. In this case the term in (\ref{H-L-phi-2}) involving $w_2$
is much smaller than the one involving $w_1$ and may be neglected.
Eq.~(\ref{H-L-phi-2}) approaches then the equation for a meson with a
single flavor \cite{KMMW} in which we replace
\beq
M^2 \to \tilde{M}^2\equiv \frac{M^2-v^2}{2} \,,
\qquad w \rightarrow w_1 \,.
\eeq
Substituting this into (\ref{massHH}) for $n=l=0$, we find
\beq
 {M}^2_{HL} = {16 w_1^2\over R^4}+{v^2\over (2\pi\alpha')^2} \label{HL-mass}
= 16 \pi {m_L^2 \over \lambda }
+{\textstyle }(m_H - m_L)^2  \,,
\eeq 
where we reintroduced the string tension $T=1/(2\pi\alpha')$
(which was set to one above) and defined the quark masses $m_{L,H}=
w_{1,2}/(2\pi \alpha')$ as the distances $w_{1,2}$ in units of
$T$.

Eq.~(\ref{HL-mass}) implies that the mass of HL mesons has
two different contributions. The first term proportional to
${m_{L}^2 \over \sqrt{\lambda}}$ has the same dependence on the 't~Hooft 
coupling as in the single flavor case \cite{KMMW}. The second term
is dominant at large 't~Hooft coupling ($\lambda \rightarrow \infty$),
where the mass of the HL mesons is approximated by the second term,
\beq \label{MHL}
 M_{HL} \approx {v\over 2\pi\alpha'} = m_H- m_L \,.
\eeq
In this strong-coupling regime, the heavy-light meson mass depends
solely on the difference of the two quark masses.  This is consistent
with the result obtained in \cite{EEG}, and provides a lower bound for the 
HL meson mass.

\begin{figure}[t] 
\begin{center}
\voffset=16cm
 \includegraphics[width=7.5cm]{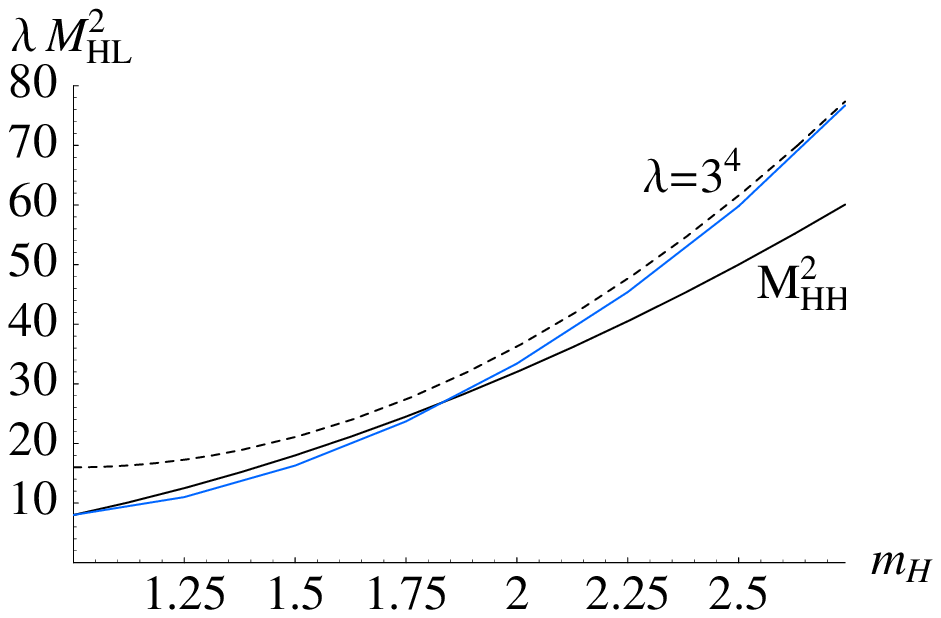}
 \includegraphics[width=7.5cm]{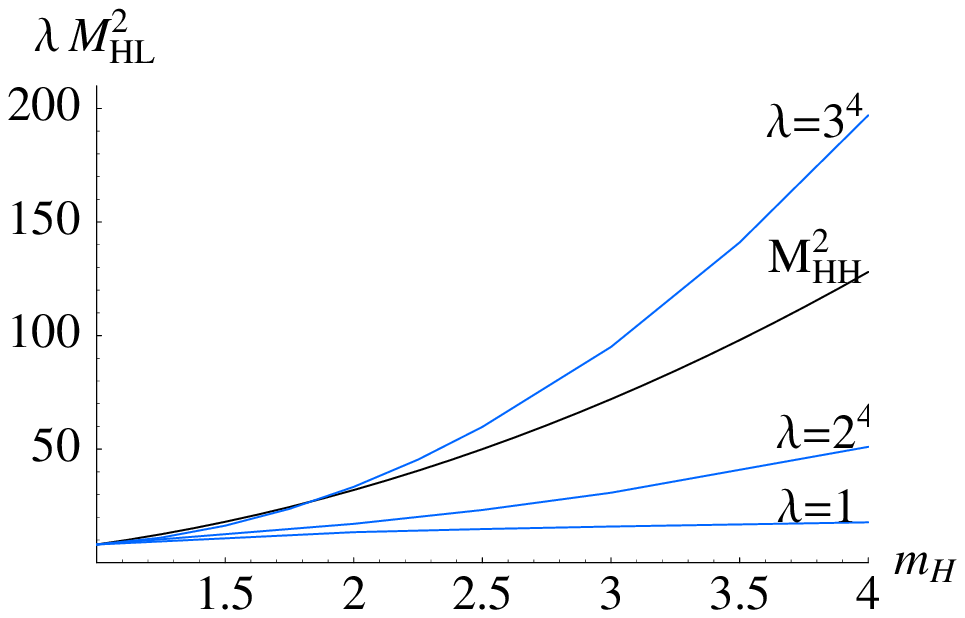}
 \caption{Numerical plots of the heavy-light meson mass for different
   values of the 't~Hooft coupling $\lambda$. Here we set $T=1$ and
   $m_L=w_1=1$.  The dashed curve shows the analytical solution
   (\ref{HL-mass}) for $\lambda=3^4$.}\label{figheavylight}
\end{center}
\end{figure}

Let us now solve Eq.~(\ref{H-L-phi-2}) numerically and compare the
results with the analytical solutions.  Fig.~\ref{figheavylight} shows
numerical plots of the heavy-light meson mass $M^2_{HL}$ (in units of
$\lambda$) versus the heavy quark mass $m_H$ for various values of the
't~Hooft coupling.  As an example, consider the asymptotics of the
graph for $\lambda=3^4$ (Fig.~\ref{figheavylight}-left). For small quark
mass differences, $m_H \approx m_L$ ($m_L=1$ in the plot), the graph
behaves very similarly to the heavy-heavy graph (\ref{massHH}) with
$m=m_H$.  For large quark mass differences, $m_H \gg m_L$, the
numerical graph asymptotes to the analytical curve given by
Eq.~(\ref{HL-mass}) (dashed curve).  We observe that the approximation
(\ref{HL-mass}) works particularly well for large masses of the heavy
quark.
 
In Fig.~\ref{figheavylight}-right we compare the heavy-light meson mass
with the heavy-heavy meson mass.  The essential difference comes again
from the 't~Hooft coupling dependence.  For small 't~Hooft coupling
the HL curve lies below the HH curve. However, there exists a critical
value of the 't~Hooft coupling for which the HL mesons are heavier
than the HH mesons. In other words, at strong coupling (at large
$\lambda$) and for fixed quark masses $m_{H,L}$, the HL meson mass is
much larger than the corresponding HH meson mass. This is of course
unphysical from the point of view of QCD. However, this seems to be a
general feature of gravity dual heavy-light models, since the 't~Hooft
coupling dependence found here coincides exactly with the one found in
\cite{EEG}.

\begin{figure}[htbp]
\begin{center}
\voffset=15cm
 \includegraphics[scale=0.73]{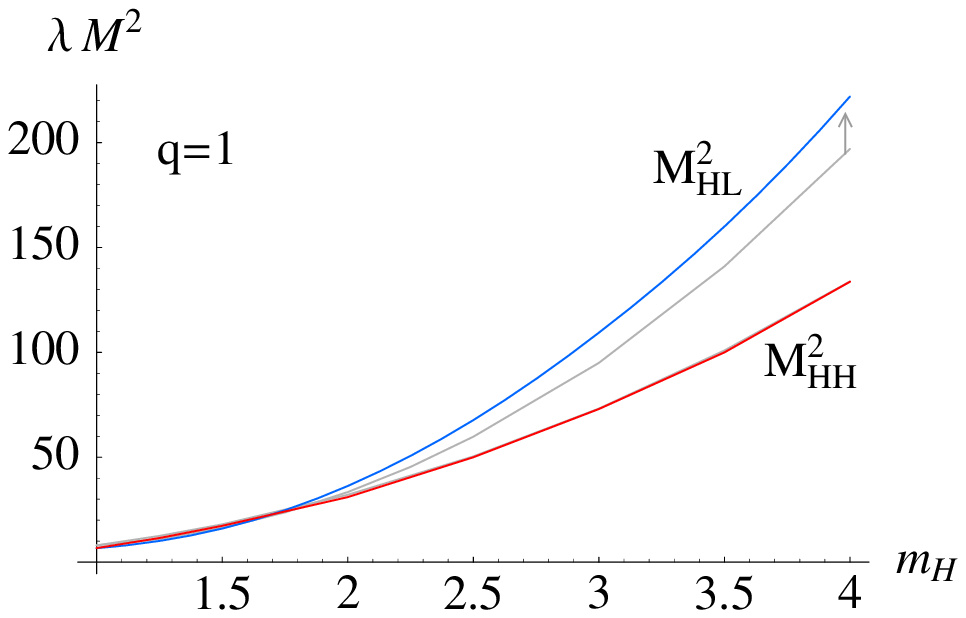}
 \includegraphics[scale=0.73]{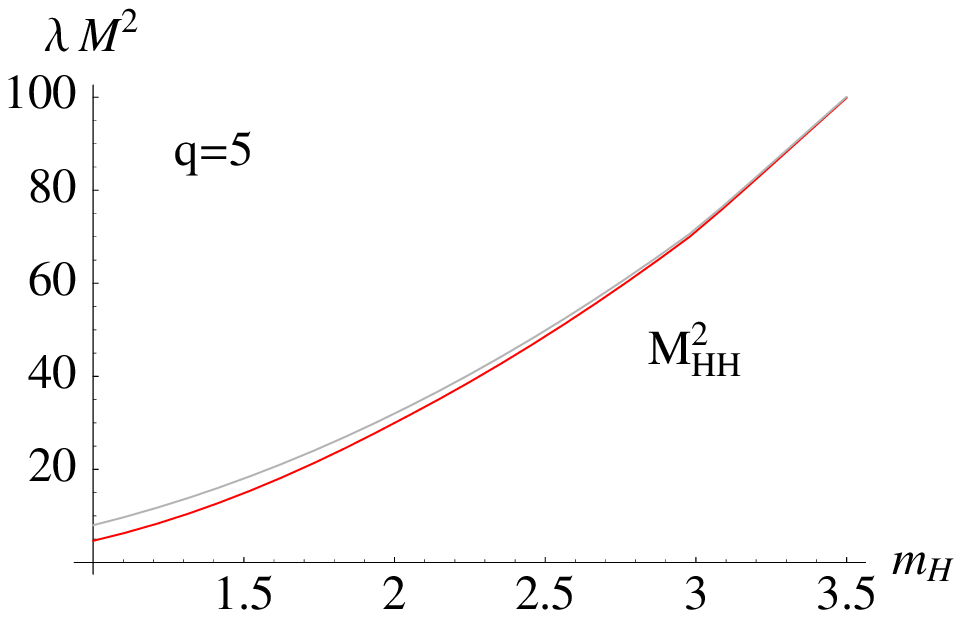}
\caption{Meson masses for non-zero $q$. The red and blue
  curves show $M_{HL}$, $M_{HH}$ for $\lambda=3^4$, $q=1$ (left) and  
  $q=5$ (right). The grey curves
  show the corresponding meson masses for $q=0$. The presence of $q$ 
  increases the HL meson masses. The lambda dependence remains unchanged.
\label{m-w2}}
\end{center}
\end{figure}

\medskip Next, as in \cite{Liu:1999fc}, we turn on a gauge condensate 
$q \sim \langle F_{\mu\nu}^2 \rangle$ such that the dual
supersymmetric gauge theory becomes confining, but remains chirally
symmetric. In general, the gauge condensate depends on the 't~Hooft
coupling and we choose $q=\bar q \lambda \alpha'^4$ as in
\cite{Liu:1999fc}.  For the dilaton given by (\ref{dilaton}),
Fig.~\ref{m-w2} shows the HL and HH mesons as a function of the heavy quark 
mass for fixed value of the light quark mass $m_L=1$. 
{We observe that the presence of $q$
increases the HL meson masses. This is due to increase of the dilaton in
the term $v^2e^{\Phi}$  in the presence of $q$. The dilaton as given by
(\ref{dilaton}) 
is responsible for quark confinement and an increase in binding energy.
Note that the term $v^2e^{\Phi}$ is independent of $\lambda$. 
On the other hand, for the HH mesons
the $q$ dependence via $e^{\Phi}$ disappears
at large quark mass $m_H$. This is seen from their equation of motion 
which is obtained from  Eq.(\ref{H-L-phi})
by replacing the last term by
$${M_{1}+M_{2}\over e^{\Phi(r_1)}+e^{\Phi(r_2)}}\to
 \partial_{\rho}\Phi(r)\partial_{\rho}
              +{M^2\over r^4}R^4 ,
$$
where
\beq
\partial_{\rho}\Phi(r)={4q\rho\over (w^2+\rho^2)(q+(w^2+\rho^2)^2)}\, .
\eeq
This shows the $q$-dependence explicitly. We see that this dependence
is small for large $w$, i.e.~for large quark mass, for all values 
of $\rho$. 
Thus in the large $w$ limit, the HH meson masses take their
${\cal N}=2$ supersymmetric value of the $q=0$ case.
This implies an increase in the difference
between the HL spectrum and the HH spectrum for sufficiently large
$\lambda$. This is
 due to the term $v^2 e^{\Phi}$ which contributes only to the HL mass.
}

\subsubsection{Vector fluctuations}

From (\ref{A2-2}), we obtain the equation of motion for the vector
fields for the confining supersymmetric background of
\cite{Liu:1999fc}.  Here we consider the four dimensional vector
$\tilde{A}_{\mu}^{1,2}$, for the case that the components
$\tilde{A}_{\rho}=\tilde{A}_{i}=0$ are zero. By imposing the gauge
condition $\partial_{\mu}\tilde{A}^{\mu}=0$, the vector fluctuation
equation of motion is
\beq
  \left(\partial_{\rho}^2+{3\over \rho}\partial_{\rho}
            -{l(l+2)\over \rho^2}+
       {M^{A}_{1}+M^{A}_{2}\over 2}\right)
          \tilde{A}_{\mu}=0 \,,  \label{H-L-A}
\eeq
\beq
    M^{A}_{i}={M^2-v^2e^{\Phi(r_i)}\over r_i^4}R^4 \,,
\eeq
where $i=1,2$ and $\tilde{A}_{\mu}^{1,2}$ are denoted by
$\tilde{A}_{\mu}$.  Note the different dilaton dependence as compared
to the scalar equation (\ref{H-L-phi-2}).  The different dilaton
dependence of vector and scalar mesons is consistent with the fact
that when adding D7 probes to the ${\rm D3}+{\rm D(-1)}$ background,
supersymmetry is broken to ${\cal N}=1$.

\begin{figure}[htbp2]
\begin{center}
\voffset=15cm
  \includegraphics[height=6.cm]{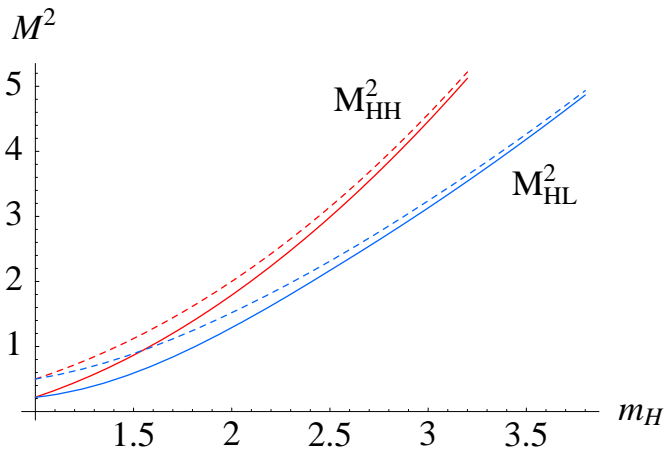}
\caption{Vector and scalar masses as function of the heavy quark mass
  for $R=2$, $q=10$. The vector masses are larger than the scalar
  masses. For large heavy quark mass, they become degenerate again. 
Red: scalar heavy-heavy meson. Red dashed: vector
heavy-heavy meson. Blue: scalar heavy-light meson. Blue dashed: vector
heavy-light meson.} \label{vectorfig}
\end{center}
\end{figure}

We find that the vector masses are slightly larger than the scalar
masses. For large heavy quark masses, the vector spectrum degenerates
with the scalar spectrum again, since the dilaton dependence becomes
negligible. For vanishing dilaton $\Phi=0$, (\ref{H-L-A}) reduces to
the same form as the scalar (\ref{H-L-phi-2}). The degeneracy of
vector and scalar masses in this limit is expected since for very
large heavy quark mass, ${\cal N}=2$ supersymmetry is restored.  This
is consistent with the phenomenological fact from heavy-quark theory
that spin effects are suppressed by powers of the inverse heavy quark
mass.  Of course, here this is due to ${\cal N}=2$ supersymmetry
restoration.  It would be interesting to compare the scalar and vector
sectors for a non-supersymmetric gravity background. However, since
the calculations are much more involved, we leave this for future
work. In the next section we consider the non-supersymmetric case for
just the scalar sector.

\subsection{Non-supersymmetric case  and chiral symmetry breaking}

The above analysis is performed for a supersymmetric background, so
$w$ is constant and there is no chiral condensate.  On the other hand,
in a background dual to a non-supersymmetric theory with chiral
symmetry breaking, the profile functions $w_{1,2}$ are not constants
but vary with $\rho$. As a result, the mass spectrum is modified due
to the presence of the chiral condensate and the related background
configuration.

We now consider the non-supersymmetric background~(\ref{dilaton-2}).
The corresponding mass spectrum of single-flavor mesons has previously
been studied in \cite{BGN}. In the following we compute the HL
spectrum dual to the scalar fluctuations $\phi^8_{1,2}$ and compare it
with the corresponding HH spectrum. We show that at strong 't~Hooft coupling
the heavy-light meson masses lie below the corresponding heavy-heavy
meson masses in agreement with phenomenological expectations.

The linearized equation of motion of the $\phi=\phi^8_{1,2}$
fluctuations is given by
\beq
  \left(\partial_{\rho}^2+{3\over \rho}\partial_{\rho}+
            {N_{\{1\}}+N_{\{2\}}\over F_{\{1\}}+F_{\{2\}}}\right)
          \phi=0  \label{H-L-phi-NS} \,,
\eeq
with
\beq
    N_{\{i\}}=\left\{F\left(\partial_{\rho}(\log F)\partial_{\rho}
              +K\right)-{v^2}G\right\}_{\{i\}} \,,
\eeq
\beq
K=(1+(\partial_{\rho}w)^2)\left(
{m^2R^4\over r^4A^4}-{l(l+2)\over \rho^2}-2K_{r}\right)\, ,
\quad K_r=\partial_{r^2}\log(e^{\Phi}A^4)\, ,
\eeq
\beq
F={e^{\Phi}A^4\over \sqrt{1+(\partial_{\rho}w)^2}}, \quad
G={e^{2\Phi}A^4 \sqrt{1+(\partial_{\rho}w)^2}}~{R^4\over r^4} \,,
\eeq
and the dilaton $\Phi(r)$ and warp factor $A(r)$ as in
(\ref{dilaton-2}). The index $\{i\}$ $(i=1,2)$ means that we have to
substitute either of the profiles $w_{1,2}(\rho)$. In the following we
consider fluctuations with quantum numbers $n=l=0$ for the sake of
simplicity.

\begin{figure}[htbp]
\begin{center}
\voffset=15cm
  \includegraphics[width=12cm]{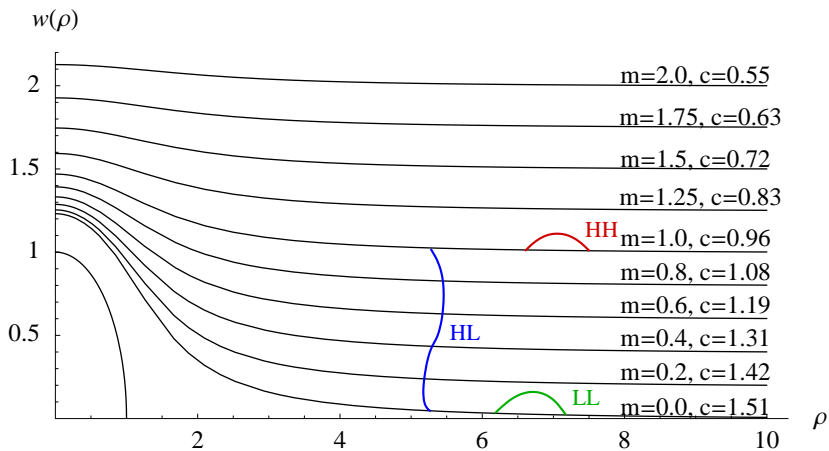}
\caption{Embedding solutions $w(\rho)$ in the nonsupersymmetric
background (\ref{dilaton-2}).} \label{embedfig}
\end{center}
\end{figure}

\vspace{.3cm} For the above non-susy configuration, the parameter
$r_0$ plays a role similar to the infra-red cut-off $\Lambda_{\rm
  QCD}$ in QCD.  Recall that the background has a singularity at
$r=r_0$ and is well-defined only for $r>r_0$ corresponding to energies
above $\Lambda_{\rm QCD}$. Quite generally, we expect the parameter
$r_0$ to depend on the (asymptotic) AdS radius $R$. For the
computation of the meson spectrum, we will make the simple choice
\begin{align}
r_0=R \,.
\end{align}
The dependence of the parameter $r_0$ on $R$ is motivated by the fact
that $r_0$ is also related to the gauge field condensate $\langle
F_{\mu\nu}^2 \rangle$, as can be seen by expanding the dilaton as
\begin{align}
e^{\phi}=1+q/r^4+\cdots \,,
\end{align}
where $q=\sqrt{6} r_0^4$. Now, $q$ is related to the gauge field
condensate $\langle F_{\mu\nu}^2 \rangle$ by $q=\lambda \langle
F_{\mu\nu}^2 \rangle$, see for instance \cite{Liu:1999fc}, and thus
$r_0 \propto R$. We should notice here that $\lambda$ is running in
the present case, since $g_{\rm YM}^2=e^{\Phi(r)}$ depends on $r$.

\vspace{.3cm} Let us now find the numerical fluctuation spectrum. As
in the single-flavor case, one first computes the D7 brane profiles
$w_{1,2}(\rho)$ by numerically solving the embedding
equation~(\ref{qeq}). The asymptotic values of $w_{1,2}(\rho)$ are
fixed by the quark masses $m_{L,H}$. The values of the quark
condensate $c=\langle \bar \psi \psi \rangle$ are obtained by
requiring the regularity of the solutions for all values of $\rho$.
For the background (\ref{dilaton-2}), the embedding solutions
$w(\rho)$ have been found in \cite{GY} and are shown in
Fig.~\ref{embedfig} for various quark masses $m$~\cite{GY}. As
in~\cite{Bab}, the solutions are repelled from the singularity at
$r_0=R=1$. Again, fluctuations of strings stretching in between two
different branes (such as the blue HL string) are dual to HL mesons.
Strings starting and ending on the same brane ({\em e.g.}~the red HH
or the green LL string) correspond to HH or LL mesons.

\begin{figure}[htbp]
\begin{center}
\voffset=15cm
  \includegraphics[width=8cm, height=5.8cm]{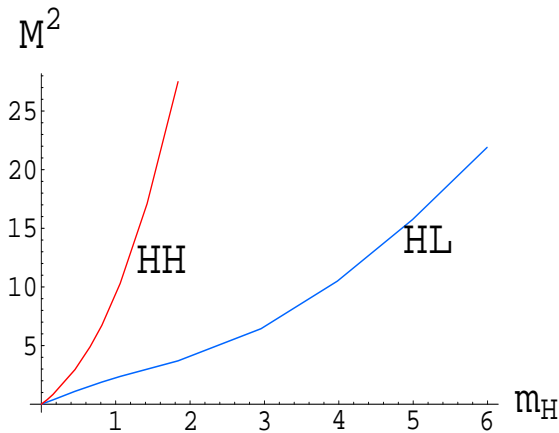}
\caption{The curves show the meson masses $M^2_{HH}$ and $M^2_{HL}$ vs
the heavy quark mass $m_H$ for zero light quark mass and $R=1$, $2 \pi
\alpha' =1$. 
\label{mHL-mq}}
\end{center}
\end{figure}

The embedding solutions must then be substituted into the equation of
motion~(\ref{H-L-phi-NS}).  Solving this equation for $\phi$, we
obtain a numerical spectrum $M(m_H,m_L, R)$ of HL mesons.
Fig.~\ref{mHL-mq} shows the resulting HL spectrum in dependence of the
heavy quark mass~$m_H$. The light quark mass is set to zero, $m_L=0$,
and $R$ is kept fixed here ($R=1$). For $m_H=0$, we recover the
(massless) Nambu-Goldstone boson in the spectrum which is expected
from spontaneous breaking of the $U(1)$ chiral symmetry~\cite{Bab,
  GY}. Both the HL and the HH spectrum satisfy the
Gell-Mann-Oakes-Renner relation $M^2 \propto m_H$ for small $m_H$. 

\begin{figure}[t]
\begin{center}
\voffset=15cm
  \includegraphics[width=7cm]{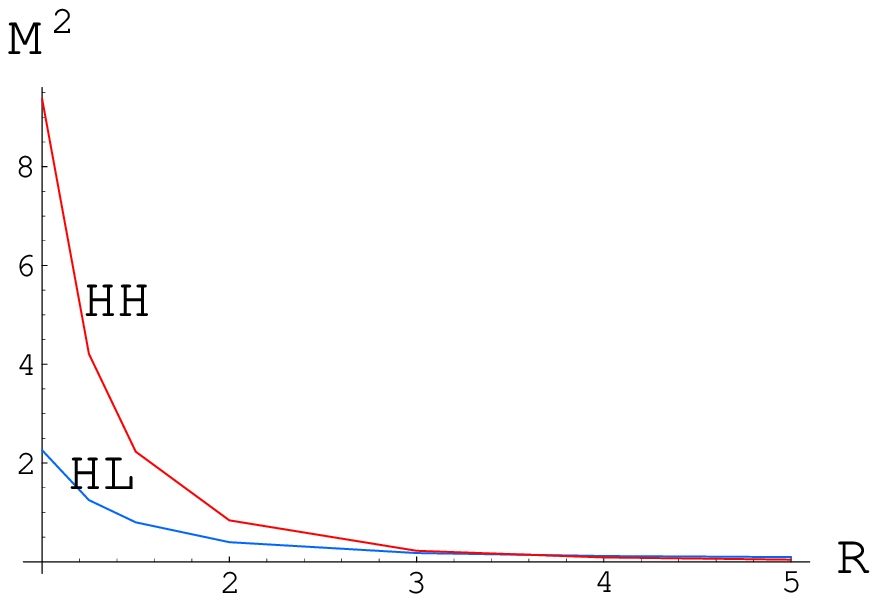}
\includegraphics[width=7cm]{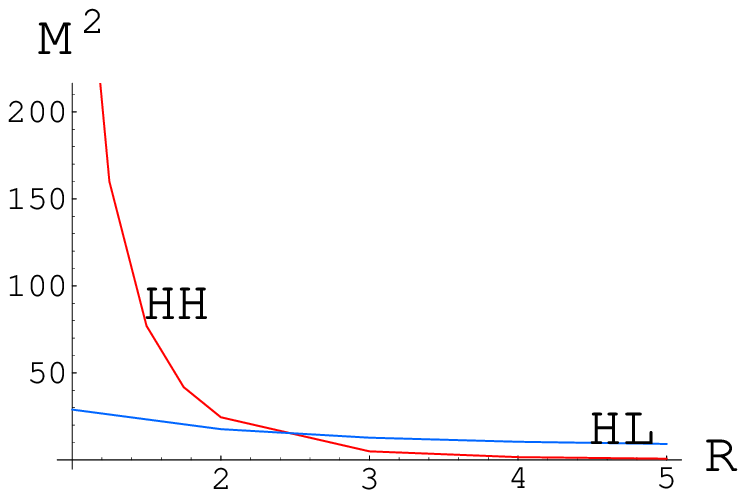}
\caption{'t~Hooft coupling $\lambda$ dependence (for $2 \pi
  \alpha'=1$, i.e.~$\lambda = \pi R^4$)
of the heavy-light and heavy-heavy mesons for the non-supersymmetric
background (\ref{dilaton-2}), for $m_H=1$ and $m_H=7$, with $m_L$=0 in
both cases. 
For large $\lambda$, the
heavy-heavy quark mass is suppressed. }
\label{nonsusylambdadep}
 
\end{center}
\end{figure}

We also find that the HL meson masses lie below the HH spectrum, at
least for an intermediate value of the 't~Hooft coupling of $1
\lesssim \lambda$, {as can be seen from Fig.~\ref{nonsusylambdadep}.
As in the supersymmetric case, the heavy-light mesons scale
differently with the 't~Hooft coupling than the heavy-heavy and
light-light ones.  There exists a critical value for the 't~Hooft
coupling (depending on $m_H$) above which the HL mesons are heavier
than the HH mesons, which is unphysical from the point of view of QCD.
Below the critical value the HH meson mass is larger than the HL meson
mass.  Accepting intermediate values of the 't~Hooft coupling in the
range of $1 \leq \lambda \lesssim 50$, it is possible to find an
appropriate parameter region where realistic mass spectra are
obtained. These spectra will be explored elsewhere.}

Moreover, we find numerically that for $r_0 \sim R \sim \lambda^{1/4}$
large, the HL spectrum $M^2(R)$ approaches $\frac{1}{2 \pi \alpha'}
(w_2-w_1)\vert_{\rho=0}$, ie.~is proportional to the distance of the
two brane probes at $\rho=0$ in the far IR.  Considering the embedding
solutions in Fig.~4, we find that this mass is equivalent to the
minimum energy of a string stretched between the two branes. At large
't~Hooft coupling, this string is much shorter than a string stretched
in between the same branes at $\rho
\rightarrow \infty$. Thus, the ${\cal O}(\lambda^0)$ contribution in the
non-supersymmetric HL spectrum is much smaller than in the
supersymmetric case, where it was proportional to $m_H-m_L$, see
Eq.~(\ref{HL-mass}).  This implies in particular that in the
non-supersymmetric case, the HL masses also tend to zero for $\lambda
\rightarrow \infty$, though more slowly than the HH masses.

\section{Wilson loop for heavy-light mesons} \label{sectionWilson}

Unlike the spectrum of single-flavor mesons, the heavy-light spectrum
of the supersymmetric theory does not vanish in the strong 't~Hooft
coupling limit, cf.\ Eq.~(\ref{MHL}). A~better understanding of this
behavior can be obtained by studying the forces (or the QCD-like tension)
between the two quarks.  For this, it is helpful to consider the quark
anti-quark potential $V_{q\bar{q}}$ which can be found by a standard
Wilson loop computation similar to that in the case of single-flavor
mesons \cite{GIN}.

\begin{figure}[t]
\begin{center}
\voffset=15cm
 \includegraphics[width=8cm]{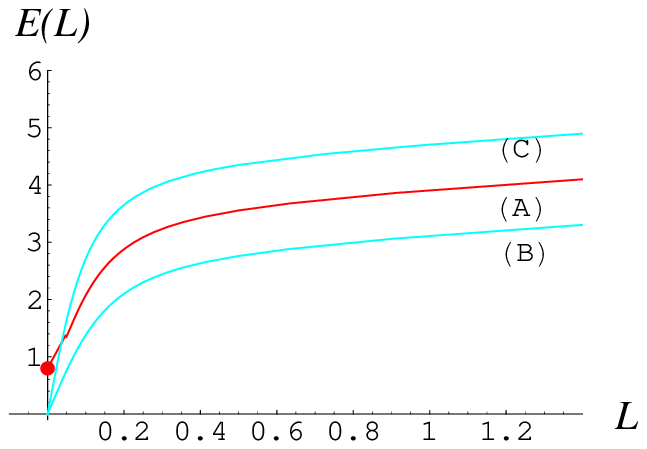}
 \includegraphics[width=7cm]{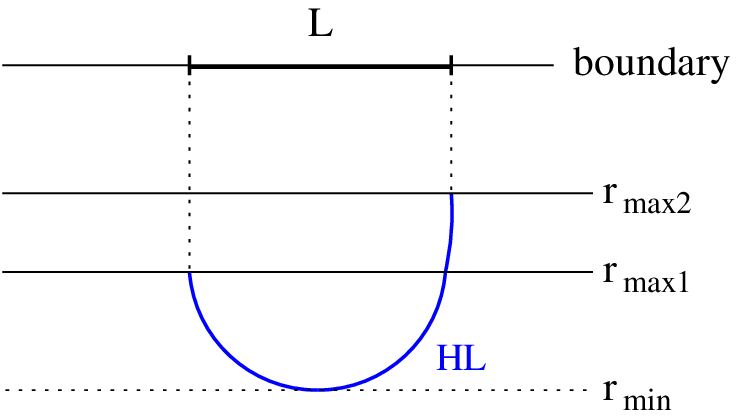}
\caption{a) Numerical plots of the energy $E(L)$ for  HL mesons (A),
 LL mesons (B) and HH mesons (C).  The circle at the endpoint of
  the curve (A) shows a finite string energy $E$ at length $L=0$.  
  Here we set $q=5$ and $R=1$, and the brane
  positions are taken at $r_{max1}=10$ and $r_{max2}=15$, 
  respectively. b) Schematic plot of the Wilson loop.
\label{Wilson-L}}
\end{center}
\end{figure}

The potential $V_{q\bar{q}}$ is derived from the expectation value of
a parallel Wilson-Polyakov loop, $ W={1\over N} \textrm{Tr} P e^{i\int
  A_0 dt}$.  In the dual gravity theory, it is represented as
\beq
    \langle W\rangle  \sim e^{-S} \,, \label{wstr}
\eeq
with Nambu-Goto action 
\beq
   S=- \frac{1}{2 \pi \alpha'} 
\int d\tau d\sigma \sqrt{-\textrm{det}\, h_{ab}} \,, 
\eeq
and induced metric
$
h_{ab}=G_{\mu\nu}\partial_a X^{\mu}\partial_b X^{\nu} 
$.   
The string world-sheet is parameterized by $\sigma$, $\tau$, which in
static gauge are set as $X^0=t=\tau$ and $X^1=x^1=\sigma$.  In the
background~(\ref{non-SUSY-sol}) the Nambu-Goto Lagrangian becomes
\beq
   {\cal L}_{\textrm{\scriptsize NG}}
   =-{1 \over 2 \pi \alpha'}\int d\sigma ~e^{\Phi/2} A(r)\sqrt{r'{}^2
    +\left({r\over R}\right)^4 A^2(r)}  \,,
 \label{ng}
\eeq
where the prime denotes the derivative with respect to $\sigma$.  We
suppose here that the test string has a deformed U-shape whose 
endpoints are on the two D7 branes, as shown in Fig.~\ref{Wilson-L}b.

\vspace{.3cm}
Let us first consider the single-flavor case. 
From the Lagrangian (\ref{ng}), we find the  relation
\beq
     e^{\Phi/2}{1\over \sqrt{(r/R)^4 A^2(r)+(dr/d{\sigma})^2}}
    \left({r\over R}\right)^4 A^3(r)= h\ ,
\eeq
where $h$ denotes a constant of motion. Here we need to introduce two
parameters, $r_{ min}$ and $r_{ max}$: $r_{ min}$ is
determined by $\partial_{{\sigma}}r|_{r_{ min}}=0$ and defines the
bottom of the deformed U-shaped string, while $r_{ max}$ is given by the
position of the D7 brane, the endpoints of the string.
Replacing $h$ by $r_{ min}$ by using the relation
$h=e^{\Phi/2}\left({r\over R}\right)^2 A^2(r)|_{r_{min}}$, we
determine the energy $E$ and the spatial distance~$L$ between the
quark and anti-quark from
\bea 
&& L=2R^2
\int_{r_{ min}}^{r_{max}} dr~ I_L ,
 \quad E=
   {1\over \pi \alpha'} \int_{r_{min}}^{r_{max}} dr~ I_E \,,
\\
  && I_E= {A(r)e^{\Phi(r)/2}\over 
     \sqrt{1-e^{\Phi(r_{min})}r_{ min}^4 A(r_{ min})^4/
             \left(e^{\Phi(r)}r^4 A(r)^4\right)}} \,,
\\
&&  I_L=  {1\over r^2 A(r)
        \sqrt{e^{\Phi(r)}r^4 A(r)^4 /
          \left(e^{\Phi(r_{min})}r_{min}^4A(r_{ min})^4\right)-1}} \,,
\eea
where $r_{min}$ is restricted as $0<r_{min}<r_{max}$. 

For the two-flavor case, $L$ and $E$ are given by
\bea
  &&  L=R^2 \left(\int_{r_{ min}}^{r_{max1}} dr~ I_L
    +\int_{r_{min}}^{r_{max2}} dr~ 
   I_L\right) \,,
\\
  && E=
   {1\over 2\pi \alpha'} 
      \left(\int_{r_{min}}^{r_{max1}} dr~ I_E
       +\int_{r_{min}}^{r_{max2}} 
       dr~ I_E \right) \,,
\eea
where $r_{max1}$ and $r_{max2}$ denote the positions of the light and
heavy quark branes, respectively. They are equivalent to $w_1$ and $w_2$
for the supersymmetric case.

For the supersymmetric background, numerical plots of the energy
$E(L)$ are shown in Fig.~\ref{Wilson-L}a.  Curve (A) is a plot of
$E(L)$ for heavy-light mesons, while (B) and (C) correspond to those
of the light-light and heavy-heavy mesons, respectively. The curves
(B) and (C) end at $(L,E)=(0,0)$; the energy is zero for vanishing
quark-anti-quark distance, $E(0)=0$. However, as mentioned above, in
the heavy-light case the energy remains finite, even for $L=0$. 

For $L=0$ the string in Fig.~\ref{Wilson-L}b stretched between the D7
branes at $r_{max1}$ and $r_{max2}$ becomes just a straight line
perpendicular to the boundary.  Then, in the Nambu-Goto Lagrangian
(\ref{ng}), we write $d\sigma=dr/r'$ and take $r' \equiv \partial r/
\partial \sigma \rightarrow \infty$. In the supersymmetric case with
$q=0$ we have $A(r)=1$, $\Phi(r)=0$. Thus, from (\ref{ng}) we get
\begin{align}
E(L=0) &= {1 \over 2 \pi \alpha'}\int dr \frac{1}{r'} \sqrt{r'{}^2
    +\left({r\over R}\right)^4 } \stackrel{r'\rightarrow \infty}{=}
\frac{1}{2 \pi \alpha'}  \, \int_{r_{max1}}^{r_{max2}} \! dr \, 
=m_H-m_L \,. 
\end{align}
This agrees with the meson mass result (\ref{MHL}). 

Note that the ordinary QCD string associated with the flux tube in
between the quark anti-quark pair can be thought of as a projection of
the Wilson loop on the boundary of the (asymptotic) AdS background,
see Fig.~\ref{Wilson-L}b. The Wilson line for $L=0$ is aligned along
the $r$-direction and therefore projected to a point on the boundary.

\section{Conclusion}

We have investigated heavy-light mesons in a holographic set-up by
considering the non-Abelian Dirac-Born-Infeld action for two D7 brane
probes embedded at different positions. The embedding matrix is chosen
to be diagonal, whereas the heavy-light mesons arise from the
off-diagonal elements of the fluctuation matrix.

We considered both supersymmetric and non-supersymmetric backgrounds
and found that the dominant contribution to the heavy-light meson mass
is ${\cal O}(1)$ in the 't~Hooft coupling $\lambda$, whereas for the
heavy-heavy and light-light mesons it is ${\cal
O}(1/\sqrt{\lambda})$. Although this result is unexpected from the
point of view of QCD, it is consistent with our Wilson loop
calculation for the heavy-light mesons, as well as with the effective
field theory holographic approach to heavy-light mesons of \cite{EEG}
which uses the Polyakov string action.
As discussed in section 4.2,
the ${\cal O}(1)$ contribution to the HL masses is proportional to the
minimal energy of a string stretching in between two branes. In the
non-supersymmetric case, the minimum energy corresponds to a string
located close to the singularity ({\em i.e.} at $\rho=0$ in Fig.~4). At
large 't~Hooft coupling this energy is much smaller than $m_H-m_L$.

As far as the spectrum of heavy-light mesons is concerned, we found that the
HL spectrum lies below the HH spectrum for intermediate values of the
't~Hooft coupling, $1 \leq \lambda \lesssim 50$. At larger 't Hooft
couplings, the HL meson masses exceed those of the HH mesons, which
would be in conflict with phenomenology. Our results are consistent
with scenarios in which QCD develops an infrared fixed point
rather than a singularity, see {\it e.g.}\ \cite{Brodsky:1997dh} and
references therein. In this case, agreement with phenomenology
would be achieved, if the gauge coupling $\alpha_s (Q^2)$ were of order 1 or
less at the fixed point (which corresponds to $\lambda = 4\pi \alpha_s
(Q^2)N_c \approx 30$).

A new element of our approach as compared to \cite{EEG} is that it allows to
distinguish scalar from vector mesons. This opens up the possibility to study
the heavy quark spin effect suppression known from QCD. We see the degeneracy
of scalar and vector masses for large heavy quark mass in the $\N=1$
supersymmetric scenario. It will be interesting to study the
non-supersymmetric case in the future.

We conclude with some remarks on the range of validity of our approach of 
using the non-Abelian DBI action for two separated branes. Generally one may 
expect
that the DBI is valid for branes separated at most by the string length $l_s$. 
Here, however, our branes are separated by a larger distance if $m_H > 1$ 
in our units of setting the string tension to one. Nevertheless our use of
the non-Abelian DBI is justified by the fact that we obtain agreement with
the classical string calculation of \cite{EEG}, as well as with the 
semiclassical Wilson loop analysis of a string stretching between the two 
branes performed in section \ref{sectionWilson}
 above in the present paper. Thus the classical
analysis at larger length scales 
dominates over quantum fluctuations at shorter scales.
This is at least in part due to the supersymmetry of the problem. We also note
that restricting to heavy quark masses of order ${\cal O (1)}$ in the string
tension will not affect the unusual $\lambda$ dependence, which intrinsically
reflects the strong coupling behaviour. Physical quark
masses correspond to $ m_H \ll 1$. In this regime, the term $m_H -m_L$ causing
the unusual $\lambda$ dependence is negligible. 
Moreover, for the non-supersymmetric approach it should be noted that the fact
that the branes approach each other in the deep interior will not alter the
$\lambda$ dependence either, since all values of the coordinate $\rho$
contribute to the action. This is again in agreement with the results of
\cite{EEG}.

\vspace{.3cm}

\section*{Acknowledgments}
K.~G.~would like to thank M.~Tachibana and Y.~Kim for discussions at
an early stage of this work. J.~E.~would like to thank Nick Evans and
Dam Son for discussions and comments. I.~K.~is grateful to Alejandro Daleo for
a discussion. The research of I.~K.~is partially supported by the
Swiss National Science Foundation and the Marie Curie network
`Constituents, Fundamental Forces and Symmetries of the Universe'
(MRTN-CT-2004-005104).


\end{document}